\documentclass[pre,aps,twocolumn,amsmath,amssymb,longbibliography]{revtex4-1}

\pdfoutput=1
\usepackage{hyperref}
\usepackage{graphicx}

\def\eqq#1{Eq.~(\ref{#1})}

\def\f#1{Fig.~\ref{#1}}
\def\fs#1{Figs.~\ref{#1}}

\def\c#1{~\cite{#1}}
\def\cc#1{Ref.~\cite{#1}}

\def\beq{\begin{equation}}
\def\eeq{\end{equation}}
\def\bea{\begin{eqnarray}}
\def\eea{\end{eqnarray}}

\def\kt{k_{\rm B}T}

\begin{document}

\title{Common physical framework explains phase behavior and dynamics of atomic, molecular and polymeric network-formers}

\author{Stephen Whitelam$^{1}$}\email{{\tt swhitelam@lbl.gov}}
\author{Isaac Tamblyn$^{2}$}\email{{\tt Isaac.Tamblyn@uoit.ca}}
\author{Thomas K. Haxton$^{1}$}
\author{Maria B. Wieland$^{3}$}
\author{Neil R. Champness$^{4}$}
\author{Juan P. Garrahan$^{3}$}
\author{ Peter H. Beton$^{3}$}\email{{\tt Peter.Beton@nottingham.ac.uk}}

\address{$^1$Molecular Foundry, Lawrence Berkeley National Laboratory, 1 Cyclotron Road, Berkeley, CA 94720, USA\\
$^2$ Department of Physics, University of Ontario Institute of Technology, Oshawa, Ontario L1H 7K4, Canada\\
$^3$School of Physics and Astronomy, University of Nottingham, Nottingham NG7 2RD, UK\\
$^4$School of Chemistry, University of Nottingham, Nottingham NG7 2RD, UK}

\begin{abstract}
We show that the self-assembly of a diverse collection of building blocks can be understood within a common physical framework. These building blocks, which form periodic honeycomb networks and nonperiodic variants thereof, range in size from atoms to micron-scale polymers, and interact through mechanisms as different as hydrogen bonds and covalent forces. A combination of statistical mechanics and quantum mechanics shows that one can capture the physics that governs assembly of these networks by resolving only the geometry and strength of building block interactions. The resulting framework reproduces a broad range of phenomena seen experimentally, including periodic and nonperiodic networks in thermal equilibrium, and nonperiodic supercooled and glassy networks away from equilibrium. Our results show how simple `design criteria' control assembly of a wide variety of networks, and suggest that kinetic trapping can be a useful way of making functional assemblies.
\end{abstract}

\maketitle

\section{Introduction}
\label{intro}

Molecular self-assembly is a promising strategy for making useful materials, and has already produced many remarkable structures in the laboratory\c{philp1996self,whitesides2002self}. But it remains largely an empirical science, in the sense that we do not know in advance which components and which conditions will give rise to successful assembly. If we could go beyond empiricism, by identifying the physical concepts and rules that underpin molecular self-assembly, then presumably we could build materials with functionalities approaching those of biological materials. The pursuit of the underlying physical principles of self-assembly motivates a large body of ongoing theoretical work -- Refs.\c{hagan2006dynamic,wilber2007reversible,rabani2003drying} being three examples -- and is the motivation for this paper. 
\begin{figure*}[ht] 
   \centering
 \includegraphics[width=0.85\linewidth]{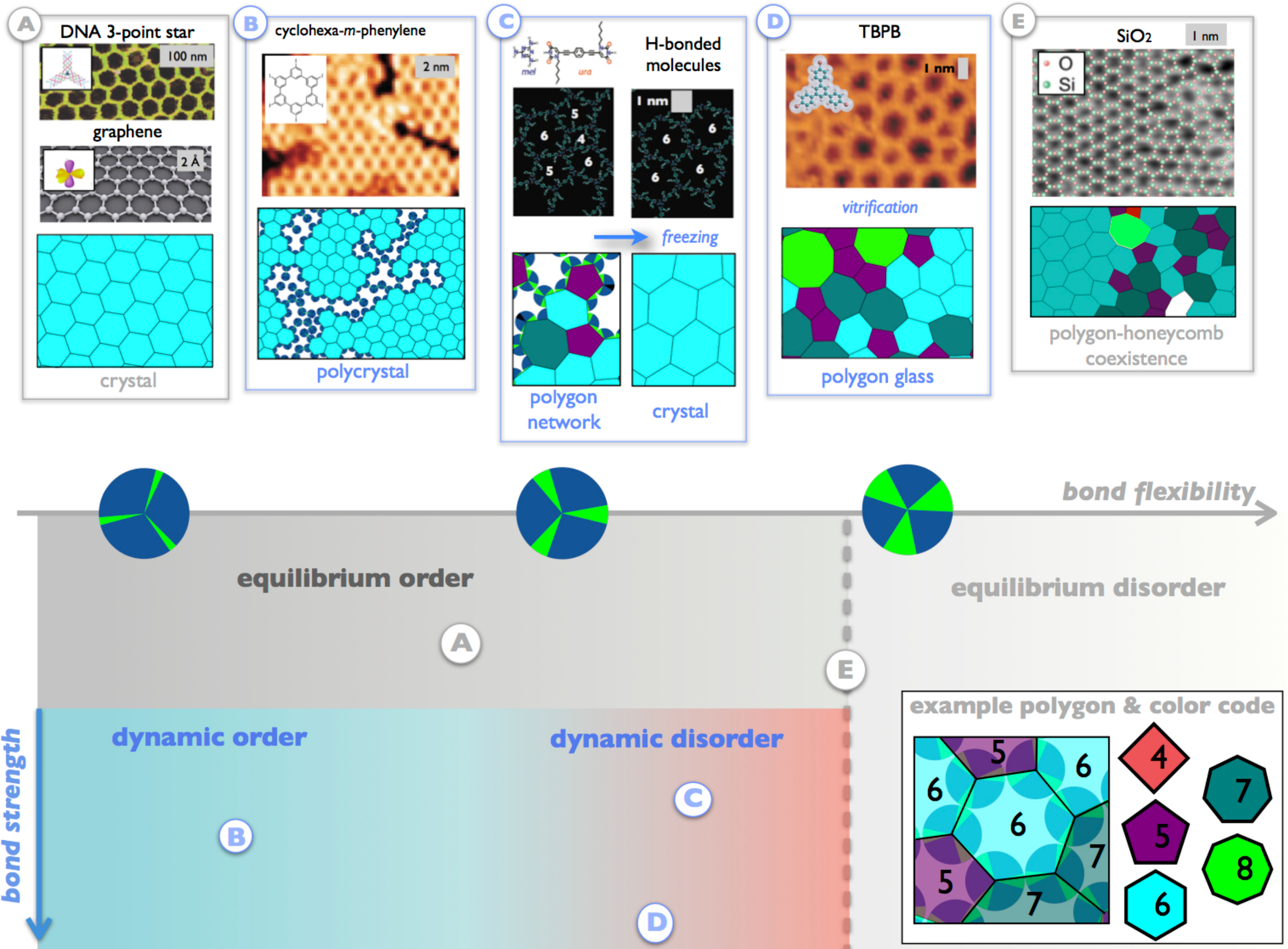} 
   \caption{{\em Spanning a lengthscale of three orders of magnitude, the networks formed by a diverse collection of building blocks can be reproduced in simulation by accounting only for the geometry and strength of building block interactions.} Three-fold-coordinated building blocks can, in equilibrium, form the periodic honeycomb network (A)\c{geim2007rise,he2005self} or a nonperiodic polygon network (E)\c{lichtenstein2012crystalline}. Dynamically, they can self-assemble as honeycomb polycrystals (B)\c{bieri2009porous}, a polygon network that evolves to the honeycomb (C)\c{palma2010atomistic}, or a kinetically trapped polygon network glass (D). Model building blocks whose interactions (parameterized by strength $\epsilon$ and flexibility $w$) are motivated by quantum mechanical calculations (\f{fig2}) can reproduce this spectrum of behavior. In equilibrium (grey lettering), such building blocks form the honeycomb network when their interactions are inflexible, and a polygon network when their interactions are flexible (\f{fig3}). Dynamically (blue lettering), within the regime of equilibrium network order, building blocks self-assemble as honeycomb polycrystals when their interactions are inflexible (few polygons are generated dynamically), and as a polygon network when their interactions are flexible (many polygons are generated dynamically). If their interactions are weak then the network evolves to the honeycomb; if their interactions are strong then the network formed is a polygon glass (\fs{fig4} and \ref{fig5}). For image permissions, please see end of paper.
\label{fig1}}
\end{figure*}

Here we take the view that in pursuit of the physical principles that underpin self-assembly there is value in identifying physical mechanisms common to apparently unlike systems. We shall show that the self-assembly of a diverse collection of building blocks, one example of which comes from our own work, can indeed be understood within a common physical framework. These building blocks range in size from atoms to micron-scale polymers made of DNA, and interact through mechanisms as different as hydrogen bonds and covalent forces. We show that in a qualitative sense the self-assembly of these building blocks, which results in a range of phenomena that include periodic and nonperiodic networks in thermal equilibrium, and nonperiodic supercooled and glassy networks away from equilibrium, can be reproduced by a statistical mechanical `patchy particle' simulation model. The model accounts {\em only} for the geometry and strength of building block interactions, indicating that these two physical factors control assembly of the real networks. Furthermore, we use quantum mechanics and analytic statistical mechanics techniques to show {\em why} we think this is so: the thermodynamics of association of model building blocks and real building blocks into isolated polygons, which one might regard as the basic constituents of self-assembled networks, is in a qualitative sense the same. This similarity reveals that the model, despite containing none of the molecular or chemical detail of the real systems, nonetheless captures a key microscopic physical feature of the self-assembly of these systems, and explains why -- or at least suggests why we should not be surprised when -- the model and real building blocks, undergoing Brownian motion, give rise to similar equilibrium and dynamic phenomena. 

In what follows we introduce the set of experimental examples we will focus on (Section \ref{examples}). We do a quantum mechanical (density functional theory, or DFT) analysis of one of these examples (Section \ref{dft}), to calculate the free energy cost of arranging molecules into isolated polygons. This calculation allows us to show that the experimental network is trapped far from equilibrium, but it also quantifies a key microscopic feature of this system, namely the thermodynamics of association of molecules into the basic polygon constituents of the network. We then introduce (Section \ref{model}) a statistical mechanical patchy particle model able to form networks. We show within a simple analytic approximation that the thermodynamics of association of model particles into polygons is similar to that of the real system studied in Section \ref{dft}. This similarity then provides a partial explanation for why equilibrium (Section \ref{thermodynamics}) and dynamic (Section \ref{dynamics}) simulations of the model reproduce the range of behavior seen experimentally. We conclude in Section \ref{conclusions}. 

In isolation, each of the techniques we have used in this paper -- self-assembly experiments, DFT calculations of assembled molecules, analytic statistical mechanical treatments of networks, and equilibrium and dynamic simulations of patchy particle models -- has been used extensively by other authors; references are given in the text. The focus of this paper is not the use of these methods individually, but the chain of connections we have drawn between experiment, the quantum mechanics of molecular interactions, and the behavior of a statistical mechanical model. We have therefore chosen to consign much of the technical detail of the individual methods to Appendices, referenced from the relevant section of text, and have focused the narrative on developing this chain of connections. Our hope is that by doing so we have written a paper that appeals to a broad readership, particularly those who are not expert with one or other of the techniques we have used.

\section{Self-assembly across scales}
\label{examples}

Let us now introduce the experimental examples on which we will focus. Panels A to E of \f{fig1} summarize a range of phase behavior and dynamics exhibited by a diverse collection of building blocks. These building blocks self-assemble into planar networks by making three pairwise bonds. When bonds are distributed regularly around the building block, the network formed is the periodic honeycomb: consider carbon atoms\c{geim2007rise} or a DNA star polymer\c{he2005self} (panel A), as well as a host of other systems\c{bartels2010tailoring}. Three-fold coordination also permits the formation of nonperiodic variants of the honeycomb. Zachariasen showed in a sketch in 1932\c{zachariasen1932atomic} that irregular 3-fold coordination results in a network of polygons of different sizes. Such a network is seen in the case of silica\c{lichtenstein2012crystalline} (panel E) on a surface. Furthermore, a range of {\em dynamics} is associated with network self-assembly. The covalently-associating molecule cyclohexa-$m$-phenylene forms polycrystals, sections of honeycomb network punctuated by grain boundaries\c{bieri2009porous} (panel B). Certain hydrogen-bonding molecules self-assemble initially as a nonperiodic polygon network that subsequently relaxes to the honeycomb\c{palma2010atomistic} (panel C). A distinct dynamics is seen in the case of the trigonal molecule tris(4-bromophenyl)benzene (TBPB) \cite{blunt2010templating} (panel D): this molecule forms a polygon network that does not evolve to the honeycomb. Preparation of this network is described in Appendix \ref{app_expt}.

\begin{figure*}[] 
   \centering
 \includegraphics[width=\linewidth]{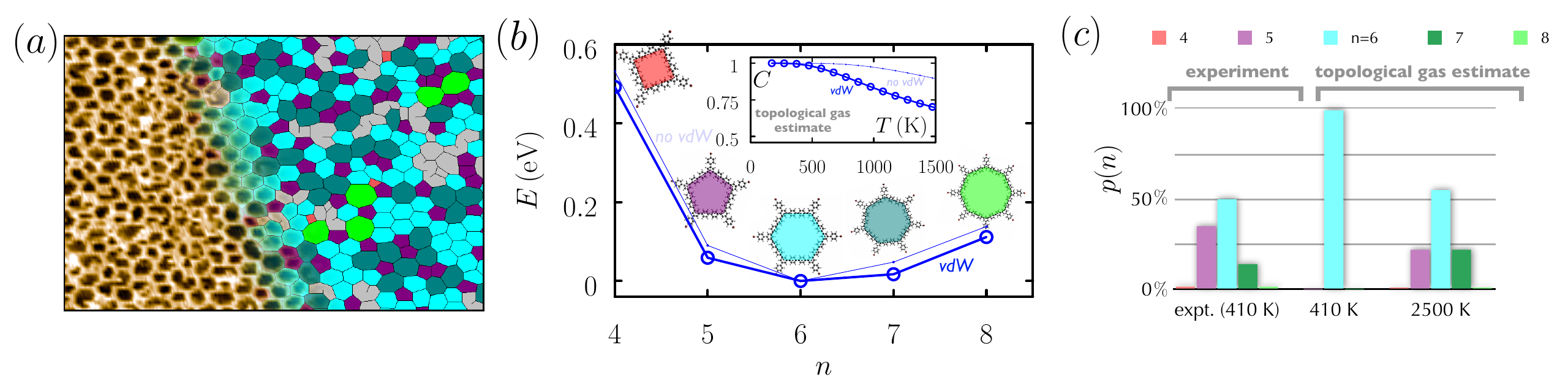} 
   \caption{{\em Analysis of one example from \f{fig1} reveals the microscopics of polygon formation.} (a) STM image of TBPB fading to polygon representation (Fig. S1). (b) DFT calculations with (vdW-DF2) and without (B3LYP) van der Waals forces show the relative energy per TBPB molecule when bound in isolated, regular $n$-gons. Using this estimate of polygon thermodynamics in a topological gas estimate (inset) shows the equilibrium network to a perfect honeycomb up to about 500 K (crystallinity $C$ is the fraction of the polygon network made up of hexagons\c{lichtenstein2012crystalline}). (c) Histogram of polygon number from experiment and as predicted in equilibrium (using the topological gas model) at two temperatures indicates that the network seen in experiment is not in equilibrium, and so is a kinetically trapped polygon glass.}
   \label{fig2}
\end{figure*}

\section{Microscopic underpinning of one particular experiment} 
\label{dft}

The spectrum of behavior seen within this class of building blocks can be reproduced within a simple physical framework that resolves only coarse details of the geometry and energetics of building block interactions (\f{fig1}, simulation snapshots and lower panel). This framework was inspired by resolving, for the particular case of TBPB, the collective microscopic mechanisms that determine the basic polygon units of the network. In \f{fig2}(a) we show a portion of the polygon network generated during TBPB self-assembly at 410 K on a gold surface (see Supplemental Information (SI)). As described in Appendix~\ref{app_dft}, we used density functional theory (DFT), using functionals with (vdW-DF2) and without (B3LYP) van der Waals interactions, to calculate the relative energy cost, per molecule, for arranging molecules into isolated, regular $n$-gons. These $n$-gons approximate the basic elements of the network. This energy cost captures the essence of the thermodynamics of molecules' polygon-forming tendencies\c{lichtenstein2012atomic,schlogl2012surface}. It is shown in \f{fig2}(b). Three features are apparent: molecules favor the hexagon, whose geometry is commensurate with the symmetry of the molecule; molecules may form other polygons, at an energy cost on a scale approaching eV (calculations done on interacting loops give similar numbers; Appendix~\ref{app_dft}); and the shape of the potential is not symmetric in $n$, as is sometimes assumed in idealized foam models\c{schliecker2002structure}.

Simple estimates based on the energy cost of forming isolated polygons of TBPB molecules suggest that the experimental network is trapped far from equilibrium. To a first approximation we see that the energy cost to turn a pair of hexagons into a heptagon and a pentagon is of order eV$/2$, indicating that in equilibrium at experimental temperatures the network should be a tiling of hexagons with characteristic linear distance between defects of order microns. As seen in \f{fig2}(a), this is not the case. At one further level of refinement, a `topological gas' calculation\c{schliecker2002structure} (see Appendix~\ref{app_top}), a mean-field thermodynamic estimate that assumes the network to be composed of isolated polygons whose average size is 6, indicates that the network in thermal equilibrium should be the honeycomb up to a temperature of at least 500 K (\f{fig2}(c)). We therefore conclude that the  polygon network seen in experiments is probably a nonequilibrium, glassy one (at this level of approximation we are not considering irregular polygons or interactions between polygons, and so we cannot prove conclusively that the network seen is a nonequilibrium one). Note that inclusion of van der Waals forces in our DFT calculations changes considerably our numerical estimate of the network ordering temperature, but not this qualitative conclusion (inset to \f{fig2}(b)). 

\section{A statistical mechanical model of network formation.} 
\label{model}

Motivated by our microscopic insight into this particular system, and by the ability of coarse-grained models to capture key physical features of complicated systems\c{ouldridge2010dna,mayer2007coarse,de2012self,hagan2006dynamic,duff2009nucleation,ten1997enhancement,rabani2003drying}, we next built a simple physical model of interacting `building blocks' in an attempt to capture the essence of TBPB's self-assembly. The model accounts only for the geometry and strength of interactions between building blocks, and pays no attention to the atomic or chemical detail through which these features arise in the real system. Although our original focus was TBPB, we found that by varying two parameters of the model -- binding strength and flexibility -- we could reproduce the behavior of {\em all} the systems described in \f{fig1}. This finding suggests that the same two factors control the self-assembly of those systems, independent of their molecular details.

\begin{figure}[] 
   \centering
 \includegraphics[width=\linewidth]{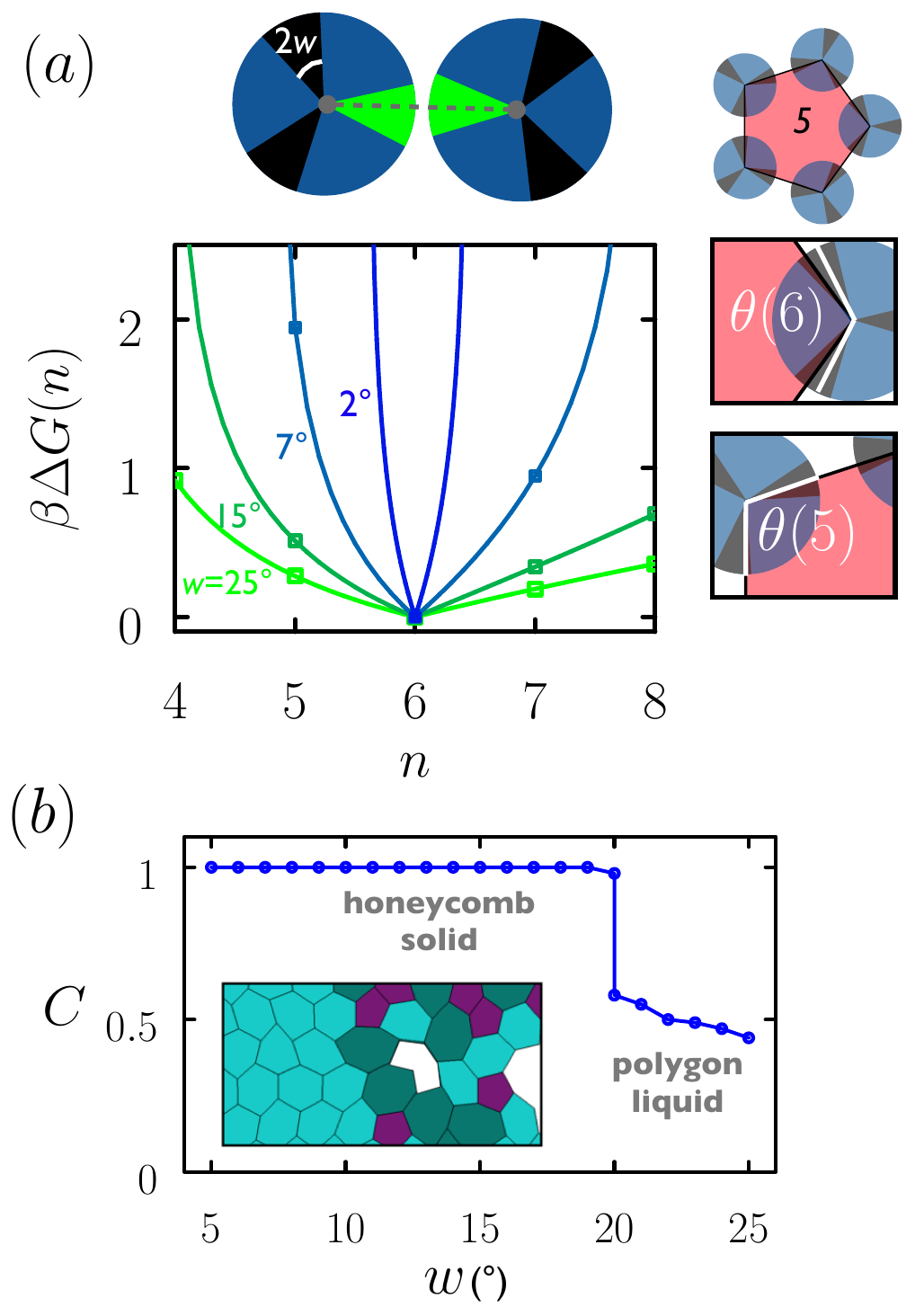} 
   \caption{{\em Model capturing the microscopics of \f{fig2} captures the range of phase behavior seen in experiments.} (a) The rotational free energy per disc within bound $n$-gons is narrow when interaction flexibility $w$ is small, and broad when $w$ is large. Although different in origin and functional form to the thermodynamics governing TBPB polygon formation, shown in \f{fig2}(b), its essential features -- the hexagon is favored, and other polygons are allowed with some geometrical strain --  are similar. Top: model geometry. Right: sketches demonstrating the geometric strain felt by discs in non-hexagonal polygons; $\theta(n) = (n-2) \pi/n$ is the internal angle of a regular $n$-gon. (b) Thermodynamic simulations ($T/\epsilon=0.16$) show that the stable network undergoes a thermodynamic order-disorder transition as a function of stripe width $w$. This thermodynamics interpolates between the examples of network order (panel A) and order-disorder coexistence (panel E) shown in \f{fig1}. Network order $C$ is the number of hexagons divided by the total number of all polygons. Inset: snapshot (Fig. S3) at thermodynamic order-disorder coexistence with $w=25^{\circ}$ (Fig. S4).}
   \label{fig3}
\end{figure}

 Following work on `patchy particle' simulation models\c{zhang2004self,tartaglia2010association,doye2007controlling,doppelbauer2010self}, we consider striped discs living on a smooth, two-dimensional substrate (Fig.~\ref{fig3}(a)). Three stripes, each of angular width $2w$, are placed regularly around the disc. Discs bind in a pairwise fashion, stripe-to-stripe\c{kern2003fluid}, with energy of interaction $-\epsilon$. Full details of the interaction potential are described in Appendix~\ref{app_model}. In figures, stripes are green when bound in this fashion. The parameter $w$ determines the flexibility of disc interactions: the broader the stripe (the larger is $w$), the less precisely need two discs align in order to bind.

When $\epsilon$ is large enough, discs can form 3-fold coordinated polygon networks. We can gain microscopic insight into the network-forming tendencies of discs by calculating the thermodynamics of isolated bound polygons of discs (the basic elements of networks), just as we did for TBPB. We calculated this thermodynamics within a simple approximation that considers only the rotational freedom discs' possess when bound in this fashion. Details of this calculation are given in Appendix~\ref{app_model}; the resulting free energy per disc as a function of polygon edge number $n$ is
\beq
\label{eq_one}
\beta \Delta G(n) =-\ln  \left( z_1(n)/z_1(6)\right),
\eeq
where $z_1(n) \equiv \max \left( 0,2w-\pi |n-6|/3 n \right)$ is the angle a disc can rotate without its stripes breaking contact with either of its two neighbors. \eqq{eq_one} is plotted in Fig.~\ref{fig3}(a). This rotational entropy is largest for the hexagon, because discs may rotate the full angular width of the stripe without breaking energetic contact. In other polygons, bound discs have less rotational freedom (as can be seen by looking at sketches of e.g. the pentagon vertex shown next to the free energy plot in \f{fig3}(a)), and so the free energy per disc is larger than in the hexagon. Rotational entropy therefore favors network {\em order}\c{mao2013entropy}. The microscopic origin of this thermodynamics (rotational entropy) is therefore different than for the TPBP molecules of Fig.~\ref{fig2} (the energy cost of irregular bond angles). {\em Despite} this microscopic difference, the essence of both systems' polygon-forming tendencies is the same: they favor hexagons, and they can achieve, with some free energy cost, other polygons. Within the model, this cost is controlled by $w$, the binding flexibility. This similarity suggests that the model, although simple, captures the physics essential to TBPB polygon formation, and, by extension, network formation (because polygons are the key constituent of the latter).

Note that the strategy of considering the free energy cost of arranging building blocks into important microscopic elements of a larger structure was used with success in\c{mao2013entropy} (compare Fig. 2(b) of that paper with our Fig.~\ref{fig3}(a)): here the same strategy allows us to compare model building blocks and real molecules in order to develop the connection between the two.
 \begin{figure*}[t!] 
   \centering
 \includegraphics[width=\linewidth]{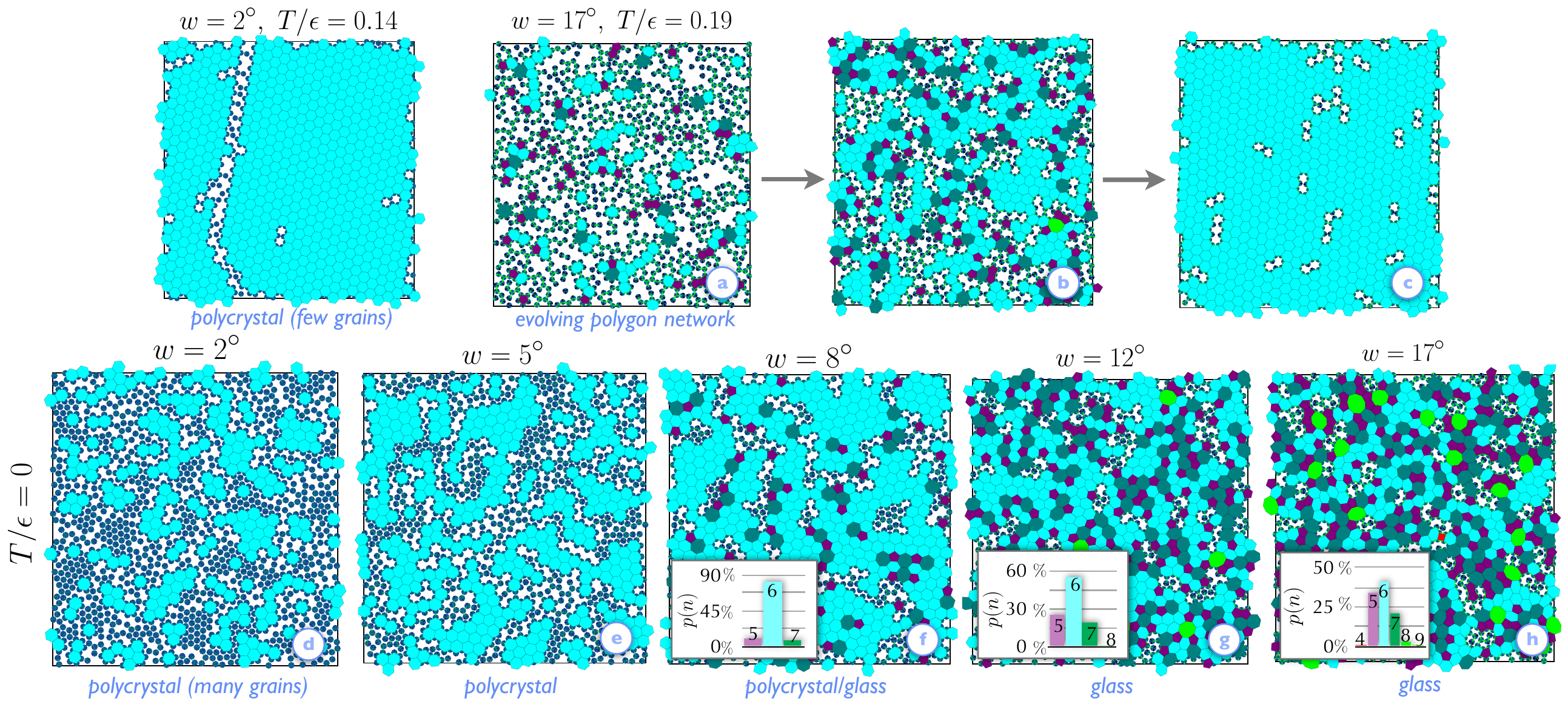} 
   \caption{{\em Snapshots of self-assembled model networks: compare the behavior of real systems in \f{fig1}.} Discs with inflexible bonds (small $w$) form polycrystals (left). Crystal grains are small if binding strength is large (bottom), similar to cyclohexa-$m$-phenylene\c{bieri2009porous}. Discs with flexible bonds (right, large $w$) form evolving polygon networks if their bonds are weak (top), similar to the hydrogen-bonding molecules of \cc{palma2010atomistic}, and form glasses if their bonds are strong (bottom), similar to TBPB (a side-by-side comparison of theory and simulation is shown in Fig. S9). Lower-case letters a--h match phase points on \f{fig5}.}
   \label{fig4}
\end{figure*}
\begin{figure*}[t!] 
   \centering
 \includegraphics[width=\linewidth]{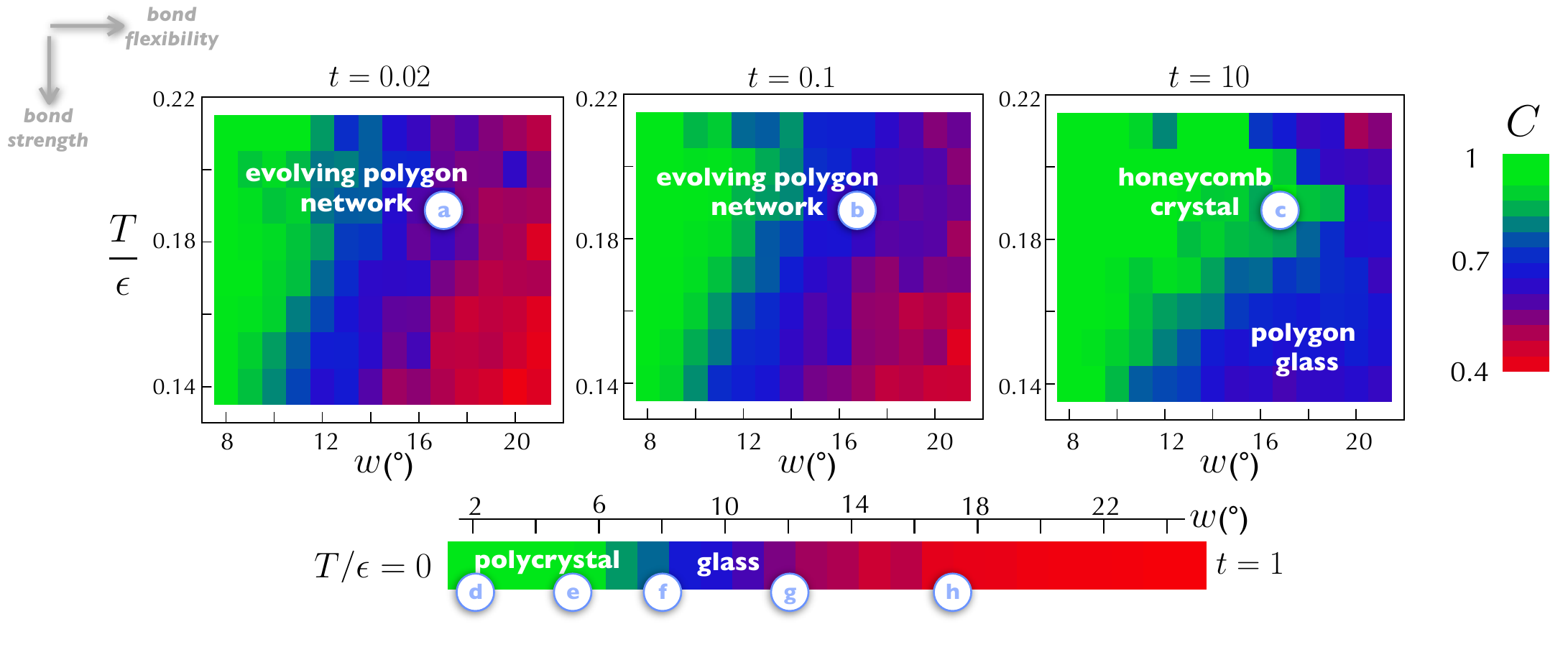} 
   \caption{{\em Model capturing the microscopics of \f{fig2} captures the range of nonequilibrium behavior seen in experiments.} We report network order $C$ (the number of hexagons divided by the total number of all polygons) in a space of inverse bond strength $T/\epsilon$ and stripe width $w$, from dynamical simulations. When $w$ is small, polycrystals assemble (see also \f{fig4}, left). For larger $w$, disordered polygon networks at early times (left panels) evolve into the stable honeycomb at later times (right panel; see also \f{fig4}, upper right), as long as bonds are weak enough to break frequently as the network assembles. Otherwise, glasses are formed. Discs with unbreakable bonds (bottom) self-assemble into structures that interpolate between polycrystals (small $w$) and glasses (large $w$). Time $t$ is measured in millions of Monte Carlo cycles. Lower-case letters a--h match snapshots on \f{fig4}.}
   \label{fig5}
\end{figure*}

\section{Model reproduces thermodynamics seen in different experiments.} 
\label{thermodynamics}

The similarity of model building blocks and TBPB molecules with respect to their thermodynamics of polygon formation leads to similar behavior in the nonequilibrium regime in which TBPB is prepared; this is described below. Moreover, by varying model parameters controlling building block binding flexibility ($w$) and strength ($\epsilon$), the model {\em also} reproduces the behavior of the other systems shown in \f{fig1}. Thermodynamically, a mean-field topological gas estimate applied to the model (details given in Appendix~\ref{app_top_model}) predicts a crossover from a honeycomb network at small $w$ (favored by discs' rotational entropy) to a polygon network at large $w$ (favored by configurational entropy). The latter is a 2D analog of a 3D patchy colloid liquid shown to be stable with respect to its crystal at zero temperature~\cite{smallenburg2013liquids}: that reference therefore identified the physics (the entropy associated with bond flexibility) that permits the fully-connected polygon network to be stable with respect to the honeycomb one.

Turning to standard equilibrium MC simulations of the discs themselves (see Appendix~\ref{app_eq}), which account for interactions and fluctuations absent from the topological gas mean-field estimate, we show in Fig.~\ref{fig3}(b) that the essence of the mean-field estimate, the change from an ordered network to a disordered one as a function of bond flexibility $w$, is confirmed by thermodynamic simulations\c{frenkel1996understanding}. (Note that in snapshots we draw polygons atop discs, but we simulated the discs themselves). In simulations, however, the transition from order to disorder is not a smooth crossover but a true phase transition. Temperature-concentration phase diagrams are shown in Fig. S4, demonstrating that in some regions of phase space there exists coexistence between ordered and disordered networks. The thermodynamics of the patchy disc model therefore interpolates between the examples of network order given in panel A of \f{fig1} (graphene and the DNA star\c{he2005self}), and the order-disorder coexistence shown in panel E of \f{fig1} (silica). This finding, combined with our analysis of the DFT results of Ref.\c{lichtenstein2012atomic} (Fig. S2), leads us to interpret the silica patterns described in Refs.\c{lichtenstein2012crystalline,lichtenstein2012atomic} as thermodynamic phase coexistence between honeycomb and polygon networks~\footnote{For work focusing on the nature of melting in 2D see e.g. Refs.\c{mehraeen2011impact,singh2013anisotropy,bernard2011two}}, albeit frozen because of the low temperatures at which images were taken. 

\section{Model also reproduces dynamics seen in different experiments.} 
\label{dynamics}

A range of nonequilibrium behavior also emerges upon variation of binding energy and flexibility. In \fs{fig4} and \ref{fig5} we report the results of dynamical simulations\c{whitelam2011approximating}, described in Appendix~\ref{app_dynamics}, in which discs were allowed to exchange with and diffuse on an initially empty substrate. When interactions are inflexible (i.e. when $w$ is small), only hexagons may form. Dynamically-generated networks in this regime are polycrystalline, having few grain boundaries in the weak bond (nucleation) regime, and many grain boundaries in the strong bond (spinodal) regime (Fig. S5). This behavior is like that of the covalent polycrystalline networks shown in panel B of \f{fig1}\c{bieri2009porous}.

By contrast, a regime in which polygons can be generated dynamically is found when building block interactions are more flexible (i.e. when $w$ is larger), still within the regime in which the network is ordered thermodynamically. Here, the initial pieces of self-assembling networks are made of a distribution of polygons, because collective microscopic motions lead to rapid formation of loops of particles that need not be six in number. When bonds are weak (i.e. when $\epsilon$ is small), this polygon network evolves to the thermodynamically stable honeycomb one. This two-step dynamics is like that seen in the H-bonded molecules shown in panel C of \f{fig1}\c{palma2010atomistic}; simulations of model clathrin honeycomb self-assembly display a similar dynamics\c{mehraeen2011impact}. When bonds are strong  (i.e. when $\epsilon$ is large), the polygon network is instead kinetically trapped, resulting in a glass. Slow relaxation of polygon defects in the face of strong bonds has been extensively discussed: see e.g. graphene\c{banhart2010structural}, clathrins\c{mehraeen2011impact} and foams\c{sherrington2002glassy}. This dynamics is similar to that displayed by TBPB, the inspiration for the model.

Glasses' polygon distributions are sensitive to rates of particle deposition, indicating that they are not simply frozen versions of the disordered network stable in equilibrium at larger $w$ (Fig. S6, Fig. S7). Instead, they are nonequilibrium structures whose polygon statistics is determined by collective microscopic motions (Fig. S8). The strong visual similarity between our simulations and experiments (Fig. S9) indicates that the model captures the physics that determines experimental patterns: molecules' substantial binding flexibility allows the formation, via a diffusive dynamics, of a range of polygons. These polygons are then `frozen in' because bonds are too strong to be broken: we calculated from DFT the bond strength of TBPB be 5 eV, an effectively unbreakable 150 $\kt$ at experimental temperatures. Our simulations also provide an explicit demonstration of the nonequilibrium origin suggested for isolated polygons made from the covalently-associating molecule 1,3,5-triiodobenzene\c{schlogl2012surface}.

\section{Conclusions.} 
\label{conclusions}

We have shown that the thermodynamic and dynamic properties of self-assembled networks whose basic lengthscales span three orders of magnitude can be reproduced within a common physical framework. This framework, developed using a combination of quantum mechanics and statistical mechanics, resolves only the geometry and strength of binding of network-forming building blocks, not their chemical and atomic details. This finding indicates that there exist basic `design criteria' -- here geometry and strength of binding -- that control the assembly of the building blocks of \f{fig1}. Our results also indicate that structure formation driven by irreversible bonds, sometimes not classed as `self-assembly'\c{whitesides2002self}, can nonetheless be considered within the same physical framework as assembly driven by reversible bonds: the behavior of covalently-associating molecules and those interacting via reversible bonds can be reproduced in different parameter regimes of the same model. The key limitation of our work is that it is of course qualitative, in respect of the comparison between experiments and statistical mechanical model. Nonetheless, quantum mechanics allows one to quantify the microscopic interactions between molecules, and so to make our approach quantitative with respect to a particular system, one could consider a statistical mechanical model with an interaction potential just complicated enough to permit exact reproduction of real molecules' free energy cost of polygon formation. We also note that we see no impediment to doing a similar study of other geometries in 2D\c{harreis2002phase}, or in 3D: indeed, recent work has shown that simplified model particles that again focus only on geometry and energy scales of binding\c{molinero2008water,saika2013understanding} (the latter being a 3D equivalent of the model studied here) can in 3D capture important structural and thermodynamic features seen in experiments done on water, and atomistic simulations of water and silica.

Our results also suggest ways of making functional materials by using kinetic trapping to generate defined nonequilibrium assemblies. Kinetic trapping, the failure of a set of self-assembling components to achieve the structure lowest in free energy, is often regarded as a nuisance, not a virtue. But the nonperiodic polygon networks studied here are generated by kinetic trapping. They have microscopic environments similar to the honeycomb, but mesoscopic environments substantially different, and so have properties not attainable to their periodic, equilibrium counterparts. Atomic-scale polygon network graphene has recently been predicted in simulations~\c{kumar2012amorphous}; this material would have novel conductance properties\c{holmstrom2011disorder}. Given that `patchy particle' models like the one use here first appeared as models of colloids, we predict that colloids -- perhaps 3-patch `lock-and-key' ones\c{sacanna2010lock} -- could self-assemble as a nonperiodic polygon network, provided that their interactions are made sufficiently strong and flexible (\f{fig5}). Such a material would have novel photonic properties\c{florescu2009designer}. \\
   
{\em Image permissions for Fig. 1}. Panel A, top, reprinted (adapted) with permission from Ref.\c{he2005self}, copyright (2005) American Chemical Society. Panel B (experimental image) reproduced from Ref.\c{bieri2009porous} with permission from The Royal Society of Chemistry. Panel C (experimental image) reprinted (adapted) with permission from Ref.\c{palma2010atomistic}, copyright (2010) American Chemical Society. Panel E (experimental image) reprinted from Ref.\c{lichtenstein2012crystalline}, copyright (2012) by The American Physical Society.

\section{Acknowledgements.} 

This work was done as part of a User project at the Molecular Foundry, Lawrence Berkeley National Laboratory, supported by the Director, Office of Science, Office of Basic Energy Sciences, of the U.S. Department of Energy under Contract No. DE-AC02--05CH11231. This research used resources of the National Energy Research Scientific Computing Center, which is supported by the Office of Science of the U.S. Department of Energy under Contract No. DE-AC02-05CH11231. IT acknowledges support from SOSCIP, NSERC, and ACEnet. MBW, NRC, JPG and PHB were supported by EPSRC Grant no.\ EP/K01773X/1. 
%
\onecolumngrid
\break
\appendix
\section{TBPB Network preparation}
\label{app_expt}
The TBPB networks were formed by subliming the molecule 1,3,5-Tris(4-bromophenyl)benzene (TBPB), which was purchased from Aldrich, onto an oriented Au(111) film grown on mica which was supplied commercially by Georg Albert Gmbh.  The experiments were performed under ultra-high vacuum (UHV) conditions in a system with base pressure $< 10^{-10}$ Torr. The Au(111) surface was first cleaned. The Au(111) surface samples were thoroughly degassed by annealing at temperatures $>600^\circ$C using a heater formed by a piece of Si(111) wafer placed behind the sample through which a current could be passed. The samples were then cleaned by repeated cycles of Argon sputtering ($\sim 5\times10^{-6}$ Torr $\sim$1.0 keV, $\sim$2.0 mA for 20 minutes) followed by annealing up to 550$^\circ$C and controlled cooling. The temperature is estimated using fixed temperature points ($T\sim 550^\circ$C, determined using a pyrometer and room temperature) and the assumption of proportionality between power output of the Si resistive heater and temperature. The TBPB is then deposited at typical rates of 1-5 monolayers/hour while heating the substrate to 100$-$150$^\circ$C. Deposition on a heated substrate is required to form the open structures discussed in the paper. Images of the resulting surfaces are acquired using a scanning tunneling microscope which operates at room temperature in constant current mode and is integrated into the UHV system. These procedures are very similar to those followed in \cc{blunt2010templating}. Processing of experimental images is described in Section S2 of the Supplemental Information.

\section{Density functional theory}
\label{app_dft}

Treatment of nanometer scale, aperiodic structures using accurate electronic structure methods is challenging. Although simulation of a single molecule of TBPB (whose chemical formula is C$_{24}$H$_{15}$Br$_{3}$) is feasible from the standpoint of computational expense, the minimum-energy, closed-loop motif includes six such units. Energetically accessible (and experimentally observed) defects comprised of eight or more units are possible. And if we consider interactions between polygons, then we must simulate larger structures still. The largest geometries we considered included over 900 electrons (398 atoms). This structure had a length of 6.5 nm along the principle axis.

To overcome these challenges, we developed a procedure based on several stages of relaxation and equilibration, each at increasing levels of theory and fidelity. Initial structures were relaxed using an interactive molecular dynamics package~\cite{avogadro} using the MMFF94 force field~\cite{mmff94}. This allowed for efficient visualization, geometry preconditioning, and motif searching.

Next, we used a minimal, localized basis set to quench the structure at the level of a hybrid-DFT functional (B3LYP~\cite{becke_jcp_1996,adamo_jcp_1997}). In the final and most computationally demanding step, we used a more complete (6-31G$^{\star \star}$) set of basis functions to completely relax the system within using the vdW-DF2~\cite{vdW-DF} framework. We use the Q-Chem code~\cite{qchem} for all of our DFT calculations. To understand and estimate effects due to dispersion interactions, we also computed energies of relaxed structures with same 6-31G$^{\star \star}$ basis set using B3LYP. From this comparison, we find that the potential energy surface predicted with vdW interactions (vdw-DF2) is more shallow and has a larger anharmonic component. B3LYP calculations are well fit by assuming the energy cost per polygon to be quadratic in the internal angle of the polygon, giving $U_{\rm B3LYP}(n) = k \left(1 - 6/n \right)^2$, with $k= 2.14$ eV. Note that this is asymmetric in $n$. vdW-DF2 calculations are fit instead by the functional form
\bea
U_{\rm vdW-DF2}(n) =  \left\{ \begin{matrix} 
     k_4  \left(\theta(n) - \theta(6) \right)^4 \qquad (4<n<8) \\ 
      \frac{1}{2} k_2 \left(\theta(n) - \theta(6) \right)^2 \qquad ({\rm otherwise}),
       \\
   \end{matrix} \right.
\eea
with $k_2 = 3.6$ eV and $k_4 = 30$ eV. A topological gas estimate (see Appendix~\ref{app_top}) allows us to compare the thermodynamics implied by the two interaction models: van der Waals forces are important quantitatively, but both functionals predict that the experimental network is glassy at 410 K.

{\em Interacting loops.} To check our understanding of this system at one further level of refinement, we performed relaxations (using the vdW-DF functional) of interacting 5-7 and 6-6 loops. Such relaxations were very costly, taking several months of computation time; we therefore used a basis set slightly smaller (6-31G$^\star$) than the one used for isolated loops. We found the 6-6 combination to be favored energetically over the 5-7 one, to the tune of 0.452 eV. Doing calculations on isolated loops using the same (slightly reduced) basis set gave a similar number, 0.435 eV, indicating that isolated-loop calculations give a reasonable representation of the behavior of molecules in connected networks.

\section{Topological gas model} 
\label{app_top}
A topological gas is a set of $M$ noninteracting $n$-gons subject to the requirement that their average size $\langle n \rangle$ is 6. This requirement comes from pretending that the $n$-gons actually form a fully connected network whose vertices are three-fold-coordinated particles\c{schliecker2002structure,schliecker2007equilibrium}. The partition function for such a gas is
\bea
Z&=& \sum_{n_1} z(n_1)  \cdots \sum_{n_M} z({n_M})  \exp\left(-\lambda\sum_i (n_i -6)\right)\nonumber \\
&\propto& \left(\sum_n z(n) \exp(-\lambda n)\right)^M,
\eea
where $z(n)$, the key input of the model, is the thermal weight of a loop of $n$ sides, and $\lambda$ is a Lagrange multiplier introduced to fix the average loop size. The loop size distribution is $p(n) = \langle{\sum_{i=1}^M \delta_{n_i,n}} \rangle$, or
\beq
p(n) = \frac{z(n) \, {\rm e}^{-n \lambda^{\star}}}{\sum_n z(n)\,  {\rm e}^{-n \lambda^{\star}}},
\eeq
with $\lambda^{\star}$ chosen to satisfy $\sum_n n p(n)=6$. The input to the model is $z(n) = \exp(-\beta U(n))$, the thermal weight of an isolated $n$-gon, which we take from DFT calculations or analytic approximations of the disc model. In the inset to \f{fig2}(b) we used as input to the topological gas model the fits $U_{\rm B3LYP}(n)$ and $U_{\rm vdW-DF2}(n)$ displayed in Appendix~\ref{app_dft}.

\section{Patchy disc model} 
\label{app_model}

Our disc model consists of hard stripy discs of diameter $a$. Discs can move in continuous space on a smooth, two-dimensional substrate. Discs are decorated by three stripes, sectors of opening angle $2w$. Stripes are arranged regularly around the disc (i.e. stripe bisectors make an angle $2\pi/3$ to the bisectors of the neighboring stripes). Discs bind in a pairwise fashion, with energy of interaction $-\epsilon$, if 1) disc centers lie within a distance $a+\Delta$, where $\Delta=a/10$, and 2) two discs' center-to-center vector cuts through one stripe on each disc (see the dotted grey line in Fig. 3(a)). This angular interaction is a 2D version of the Kern-Frenkel potential\c{kern2003fluid}. To ensure that a stripe can bind to only one other stripe, we restricted the patch opening angle to $w<\arcsin\left(\frac{a/2}{a+\Delta}\right)= \arcsin(5/11) \approx 27.0^{\circ}$.

{\em Polygon-forming thermodynamics of the disc model.} We can estimate the free energy cost of an isolated regular $n$-gon, the objects considered in our DFT study of TBPB, by considering the angle each disc in a regular $n$-gon may rotate while its two stripes are bound to stripes on neighboring discs (see \cc{mao2013entropy} for an elegant general theory accounting for rotational entropy in periodic assemblies). We assume particle centers to be fixed (i.e. we neglect vibrational entropy). To estimate rotational entropy, we note that each internal angle of a regular $n$-gon is $\theta(n)=(n-2)\pi/n$, while $\theta(6) = 2 \pi/3$ is the angle between adjacent stripes on a disc. The angle $z_1(n)$ a disc can rotate without its stripes breaking contact with either of two neighbors in an $n$-gon is its stripe width $2w$ minus the (magnitude of) the difference between $\theta(n)$ and $\theta(6)$, i.e. $z_1(n) = \max\left(0,2w - |\theta(6)-\theta(n)|\right)$. This can be written
\beq
\label{pot}
z_1(n)=\max \left( 0,2w-\frac{\pi}{3 n} |n-6| \right).
\eeq
This angle is largest for the hexagon, where it is equal to $2 w$, the width of the patch. To this level of approximation, the thermal weight of an $n$-gon is $z(n)=z_1(n)^n$. Rotational entropy therefore favors network order (networks made of hexagons). In Fig. 3(a) we plot for different choices of $w$ the (normalized) free energy per disc associated with this rotational partition function, namely $\beta \Delta G(n) =-\ln  \left( z_1(n)/z_1(6)\right)$.

\section{Topological gas estimate applied to patchy disc model} 
\label{app_top_model}

We can get a rough sense for how the thermodynamics of the disc model network depends on bond flexibility by using the polygon free energy cost, \eqq{eq_one}, as the input $\beta \epsilon(n)$ to the topological gas model. We have
\beq
\label{top}
p(n) =\frac{ \exp\left( -\lambda^{\star} n -\beta \epsilon(n) \right)}{\sum_n \exp\left( -\lambda^{\star} n -\beta \epsilon(n)\right)},
\eeq
where 
\beq
\label{internal_nrg}
\epsilon(n) = -n \, \kt \ln \max  \left( 0,2w-\frac{\pi}{3 n} |n-6| \right)
\eeq
is the free energy cost of a loop of $n$ sides (here, for simplicity, we ignore the possibility of broken bonds and network compressibility, although both effects arise in simulations of discs). 

The Helmholtz free energy per loop of the network is $f_{\rm net}= -TS+U$ or
\beq
\label{network}
f_{\rm net}=\kt \sum_n p(n) \ln p(n) + \sum_n p(n) \left[\epsilon(n) +\lambda^{\star} n \right].
\eeq
The first term in \eqq{network} is $-T$ times the configurational entropy of the network loop distribution. This entropy favors network {\em disorder:} it is large for a broad distribution of loop sizes, and zero for the pure honeycomb network (for which $p(n) =\delta_{n,6}$). The second term contains the `internal energy' $\epsilon(n)$ of each loop, \eqq{internal_nrg}. For the disc model this is entropic in origin, and comes from the rotational entropy of particles in the loop. It is largest for the honeycomb network, and so this entropy favors network {\em order}. The piece $\lambda^{\star} n$ enforces the Euler constraint that the average loop size is 6, and can be regarded as an effective loop chemical potential. \eqq{top} predicts a smooth crossover from an ordered network to a disordered one beginning at a patch width of about $w=10^{\circ}$. Simulations (Fig. 3(b)) show instead an order-disorder phase transition closer to $w=20^{\circ}$. This numerical difference is expected because 1) our analytic estimate for discs' polygon-forming thermodynamics ignores vibrational entropy, and 2) the topological gas approximation we have used ignores polygon-polygon interactions, and hence surface tension (note though that polygon interactions can be included within a topological gas framework\c{schliecker2002structure}). Nonetheless, analytic study of the disc model identifies the physics responsible for the order-disorder phase transition seen in equilibrium simulations.
 
\section{Equilibrium simulations of the disc model}
\label{app_eq}

We calculated network thermodynamics in Fig. 3(b) by performing direct coexistence simulations, Gibbs ensemble simulations, and fixed-pressure Monte Carlo simulations\c{frenkel1996understanding}, in all cases using approximately 1000 discs per simulation. Fig. S4 shows two characteristic phase diagrams in the conventional temperature-density plane.

For small widths (e.g. $w=10^\circ$), there are only two coexisting phases: a monomer fluid at low density, and a solid at high density.  We calculated the properties of these phases by equilibrating a solid slab set in contact with a fluid slab, within a periodic rectangular box.  We set the size and initial shape of the box so that approximately $75\%$ of the discs would be in an approximately square-shaped solid slab.  We allowed the box lengths to fluctuate at constant area to equilibrate the stress.  At low temperatures, we found that the solid phase is a honeycomb network with a packing fraction $\phi\simeq 0.55$.  At high temperatures, the solid phase becomes partially filled with discs at the interstices of the honeycomb network. This filling is shown by the change in density in Fig. S4 (a), signaling a crossover toward a hexagonal phase at high temperature.  We checked that the properties of the coexisting phases were the same regardless of how the solid slab was initialized (as a honeycomb, hexagonal phase, or partially-filled honeycomb), and the same regardless of how the gas slab was initialized (as a vacuum phase or a high-temperature gas).  At high temperatures, we could only compare the last two initial conditions, because the coexisting fluid became denser than the honeycomb; even a box filled with honeycomb would melt into a single-phase gas.

For larger widths (e.g. $w=25^\circ$; see Fig. S4(b)) a polygon liquid phase emerges at intermediate temperatures.  We simulated these phases in the Gibbs ensemble~\cite{Panagiotopoulos1989}.  As shown in Fig. S4 (b), we found that we could fit a binodal of the form expected for the two-dimensional Ising universality class,
\begin{equation}
(\phi_{\rm liquid}-\phi_{\rm gas})^8=c_1(T_{\rm c}-T),
\label{Ising}
\end{equation}
\begin{equation}
\dfrac{1}{2}\left(\phi_{\rm liquid}+\phi_{\rm gas}\right)=\phi_{\rm c}+c_2(T_{\rm c}-T),
\label{rectilinear}
\end{equation}
where $(\phi_{\rm c}, T_{\rm c})$ is the critical point, $c_1$ and $c_2$ are constants, and \eqq{rectilinear} is the empirical law of rectilinear diameter.  As for smaller $w$, we obtained gas-solid coexistence densities using direct coexistence simulations, finding that the solid is a honeycomb network.  Although we could use direct coexistence simulations to observe polygon liquid-honeycomb solid coexistence above the triple point, the interfaces between the slabs were not stable enough to accurately calculate the properties of the coexisting phases.  We expect that to due to low interfacial tension between the phases, such direct coexistence simulations would have to be conducted with much larger systems.  Instead, we estimated the properties of the coexisting liquid and solid phases by performing fixed-pressure simulations at a range of pressures.  Since we initialized the simulations in the solid phase, we characterized the coexisting liquid as the highest-pressure system that melted and the solid as the lowest-pressure system that remained a stable solid, using pressure steps of size $0.2 \epsilon a^2$, where $a$ is the disc diameter.  Depending on temperature, we initialized the systems either as the honeycomb or as hexagonal crystals.  Starting from a strongly unstable crystal (hexagonal at low $T$ or honeycomb at high $T$) led to prohibitively slow equilibration.  As shown in Fig. S4 (b) for $w=25^\circ$, the solid phase crosses over from a honeycomb crystal with $\phi\simeq 0.56$ to a hexagonal crystal with $\phi\simeq 0.80$ as temperature increases.
 
In Fig. 3(b) we define the network order at coexistence as the network order of the first condensed phase upon compression.  Choosing $T/\epsilon=0.16$, the first condensed phase is the honeycomb solid for $w\le 20^\circ$ and the polygon liquid for $w\ge 21^\circ$.

\section{Dynamical simulations of the disc model} 
\label{app_dynamics}

\f{fig4} and \f{fig5} were obtained from dynamical simulations of the following nature.  The substrate was initially empty. Discs were allowed to bind to or unbind from the substrate (assuming an implicit solution of discs in contact with the substrate), and to translate and rotate diffusively on the substrate. To approximate on-substrate diffusive motion we used the virtual-move Monte Carlo algorithm\c{whitelam2007auk,whitelam2011approximating}. This algorithm moves particles locally according to gradients of potential energy, and collectively so as to approximate diffusion expected of overdamped motion. We checked that conventional single-particle moves reproduce (in a qualitative sense) the classes of structures -- polycrystals, glasses etc. -- described in the text. We expect therefore that our qualitative conclusions are likely to be independent of precise details of the dynamic protocol used.

To move particles to and from the substrate we used grand canonical Monte Carlo moves, namely single-particle insertions (proposed anywhere in the box) and deletions (of randomly-chosen single particles), proposed with equal likelihood. The acceptance rate ratio for these moves is\c{frenkel1996understanding}
\beq
\label{ratio}
\frac{p_{\rm acc}(N \to N+1)}{p_{\rm acc} (N+1 \to N)} =  \frac{p_{\rm prop}(N+1 \to N)}{p_{\rm prop}( N \to N +1)} \frac{V}{N+1}  {\rm e}^{ \beta \mu - \beta \Delta E },
\eeq
where $V$ is the box volume, $\Delta E$ is the energy change resulting from the proposed move, $p_{\rm prop}(N \to N+1)$ is the rate at which the insertion move is proposed, and $p_{\rm prop}(N+1 \to N)$ is the rate at which the deletion move is proposed.

Choosing grand canonical and on-substrate moves with fixed probabilities (Method 1) gives $p_{\rm prop}(N \to N+1)=p_{\rm prop}(N+1 \to N)$, and so appropriate choices for insertion and deletion acceptance rates are
\beq
\label{rate1}
p_{\rm acc}(N \to N+1) =\min \left(1, \frac{V}{N+1}  {\rm e}^{ \beta \mu - \beta \Delta E } \right)
\eeq
and 
\beq
\label{rate2}
p_{\rm acc}(N \to N-1) = \min \left(1, \frac{N}{V} {\rm e}^{ -\beta \mu - \beta \Delta E }\right).
\eeq
 
However, choosing diffusion and grand-canonical moves with fixed probabilities results (particularly at low temperature, where bound discs rarely unbind) in a dynamics in which the effective on-substrate basic diffusion rate becomes more sluggish as the substrate becomes host to more particles. To illustrate this effect, consider the case in which diffusion and grand-canonical moves are chosen with equal likelihood. If one particle is present on the substrate, then its frequency of motion with respect to that of particle deposition is 2:1. But if 100 particles lie on the substrate, the frequency of motion of each, relative to that of particle deposition on the substrate, is 1:50.

To counter this effect, we also did simulations (Method 2) in which grand canonical moves were proposed with likelihood $1/(N+1)$, where $N$ is the instantaneous number of particles on the substrate. In this case, $p_{\rm prop}(N \to N+1) \propto 1/(N+1)$, and $p_{\rm prop}(N+1 \to N) \propto 1/(N+2)$ (with the same constant of proportionality, $1/2$). From \eqq{ratio} it can be seen that in order to preserve detailed balance, the replacement $N \to N+1$ must be made to the right-hand sides of the acceptance rates~\eqq{rate1} and \eqq{rate2}. The relative proposal rate of on-substrate diffusion and particle addition is then independent of $N$. No modification of the on-substrate move acceptance rates is needed: if $N$ particles lie on the substrate, then both forward and reverse diffusive moves are proposed with rate $N/(N+1)$, and so this factor cancels from the detailed balance condition for those moves. This dynamics is still approximate, in a physical sense, because it assumes that removal of discs from the substrate is not important (the rate for this process could in principle be scaled independent of the deposition rate, but we have chosen not to do this). Simulation results in the text are from Method 2, but those from Method 1 are qualitatively similar (meaning that the regimes of parameter space in which we see polycrystals, glasses, evolving polygon networks etc. are similar). Larger differences were observed within each method by varying the relative rates of deposition (see Figs. \ref{fig4} and \ref{fig5}).

The chemical potential $\mu$ was chosen so that the disc packing fraction in the absence of attractive interactions was 25\%. In Figs. S6 and S7, the case of `fast deposition' corresponds the procedure just outlined, while `slow deposition' corresponds to a similar dynamics in which the basic rate of grand canonical moves was reduced by two orders of magnitude.

\end{document}